\documentclass[twocolumn]{aastex631}

\usepackage{hyperref}
\usepackage{soul}

\newcommand{\chandra}{\textit{Chandra}}

\newcommand{\xmm}{\textit{XMM-NEWTON}}

\newcommand{\ms}{\ensuremath{M_{\odot}}}

\newcommand{\lumcgs}{\ensuremath{\mathrm{erg}\,\mathrm{s}^{-1}}}

\shorttitle{Hard X-ray flares and spectral variability in NGC 4395 ULX1}
\shortauthors{Ghosh et al.}
\graphicspath{{./}{figures/}}

\begin{document}

\title{Hard X-ray flares and spectral variability in NGC 4395 ULX1}

\correspondingauthor{Tanuman Ghosh}
\email{tanuman@rri.res.in}

\author{Tanuman Ghosh}
\affiliation{Department of Astronomy \& Astrophysics, Raman Research Institute, C. V. Raman Avenue, Sadashivanagar, Bangalore 560080, India}

\author{Vikram Rana}
\affiliation{Department of Astronomy \& Astrophysics, Raman Research Institute, C. V. Raman Avenue, Sadashivanagar, Bangalore 560080, India}

\author{Matteo Bachetti}
\affiliation{Istituto Nazionale di Astrofisica-Osservatorio Astronomico di Cagliari, via della Scienza 5, I-09047 Selargius (CA), Italy}




\begin{abstract}
We report the detection of flaring events in NGC 4395 ULX1, a nearby ultraluminous X-ray source (ULX), for the first time, using recent \xmm\ observations. The flaring episodes are spectrally harder than the steady emission intervals, resulting in higher fractional variability in the high energy regime. A thin Keplerian and a slim accretion disk provide the best-fit continuum for \xmm\ spectra. All observations show a broad hump-like feature around $\sim 0.9$ keV, which can be associated with a collection of blended emission lines, and suggests the presence of a wind/outflow in this ULX through comparison with other ULXs that show a similar feature. The flaring spectra correspond to higher slim disk temperatures due to higher mass accretion rate under an advection-dominated accretion scenario. The luminosity-temperature (L-T) values in different flux states show a positive trend. When characterized with a powerlaw relation, the L-T profile is broadly consistent with both $L\propto T^2$ and $L\propto T^4$ relations for the analysed data. The empirical predictions for a slim accretion disk in the case of super-Eddington accretion onto a stellar-mass compact object is $L \propto T^2$ which is a possible scenario in ULX1. The origin of the flaring events is understood as an intrinsic change of accretion rate or presence of variable clumpy wind in the inner region of the accretion disk.
\end{abstract}



\section{Introduction} \label{sec:intro}
Ultraluminous X-ray sources are the brightest off-nuclear X-ray binaries, having emission luminosities above the Eddington limit of a 10 \ms  black hole (L$_x > 10^{39}$ \lumcgs; see \citealt{Kaaret2017} for a recent review). Compared to the sub-Eddington Galactic X-ray binaries (XRBs) or active galactic nuclei (AGNs), ULXs show distinct spectral curvature below 10 keV (see for example \citealt{Bachetti2013,Walton2013,Walton2014, Walton2015,Walton2015J,Rana2015,Mukherjee2015,Furst2017}), suggesting that these sources are mostly super-Eddington stellar-mass compact object accretors. 
Indeed a number of ULXs were found to have neutron star accretors, confirming this super-Eddington interpretation \citep{Bachetti2014, Furst2016N,Israel2017F,Israel2017M,Brightman2018,Wilson-Hodge2018,Vasilopoulos2020, Chandra2020,Carpano2018,Rodriguez2020,Sathyaprakash2019}.

ULX spectra show two common features - a characteristic turnover below 10 keV and a soft excess $\leq 1$ keV. These properties are typically explained by disk wind emission in the scenario of super-Eddington accretion \citep{Shakura1973, Poutanen2007}. The mass outflow rate and the viewing angle of the disk determine the spectral softness, with softer sources being observed nearer to the plane of the disk (see \citealt{Middleton2015Mar,Pinto2021} and references therein). Strong blue-shifted atomic features are one of the signature characteristics of such relativistic ($\beta \sim 0.25$) wind emission \citep{Pinto2016,Pinto2017,Kosec2018a,Kosec2018b,Walton2016,Middleton2015Dec,Pinto2020} which can share a large fraction of the total energetic budget of ULXs.

ULXs are broadly categorized into four classes \citep{Sutton2013,Kaaret2017}. ``Broadened disk" (BD) sources exhibit hot thermal spectra from a geometrically modified disk, related to the super-critical slim disk scenario \citep{Begelman1979,Abramowicz1988,Abramowicz1989}. These sources belong to the lowest luminosity regime of ULXs ($\sim 1-3 \times 10^{39}$ \lumcgs), and the typical accretion rate of these sources is near or slightly above the Eddington accretion rate. The higher luminosity ULXs mostly show two-component spectra in $0.3-10.0$ keV energy range, a soft thermal component, and a hard component. Depending on the hardness of the sources, they are classified as ``Hard Ultraluminous" (HUL) or ``Soft Ultraluminous" (SUL) sources.

ULXs with powerlaw index $\Gamma > 2$ in $0.3-5.0$ keV energy band are classified as SUL sources \citep{Pinto2021}. Among them, a sub-class is ``supersoft ultraluminous" (SSUL) sources or ``Ultraluminous supersoft sources" (ULSs), which have most of their observed flux at energies below $\sim 1$ keV. The bolometric luminosity of these sources is typically a few $10^{39}$ \lumcgs and their spectra are mostly dominated by cool blackbody component ($T_{bb}< 0.14$ keV; \citealt{Urquhart2016,Pinto2021}). However, there exists some sources, like NGC 55 ULX and NGC 247 ULX-1, which are considered to be soft ULX sources but fall somewhere between ULX and ULS spectral states (See figure 1 of \citealt{Pinto2017}). Unlike most ULSs, they show spectra that extend up to $\sim 5$ keV, and typically, their spectra are best modeled by two blackbody components \citep{Pinto2017,Pinto2020,Pinto2021}.

Modern X-ray instruments have observed a large number of ULXs. However, few sources have shown short-term timing variability in terms of fractional variability, quasi-periodic or periodic oscillation. A number of sources like NGC 1313 X1 \citep{Walton2020}, NGC 7456 ULX-1 \citep{Pintore2020}, NGC 253 ULX-1 \citep{Barnard2010}, NGC 6946 ULX-3 \citep{Earnshaw2019}, NGC 247 ULX-1 \citep{Pinto2021}, 4XMM J111816.0-324910 in NGC 3621 \citep{Motta2020}, NGC 4559 X7 \citep{Pintore2021}, M82 X-1 \citep{Brightman2016} have been found showing some intermittent flaring events. These flaring activities can help shed light on the dynamics of accretion processes or wind outflows in these sources.

NGC 4395 ULX1 (2XMM J122601.4+333131; \citealt{Liu2005}) is a ULX which has shown a long term variability \citep{kaaret2009}. \citealt{Vinokurov2016} suggested that the source possibly exhibited a period of 62.8 days in archival observations. \citealt{Earnshaw2017} studied this source using previous \xmm\ and \chandra\ observations. Those data showed no significant short-term timing variability in this source. In this paper, we present a detailed study of four high-quality \xmm\ observations, two of which show flaring activities from this source for the first time.

\begin{deluxetable*}{ccccc}
\tablenum{1}
\tablecaption{Observation log of NGC 4395 ULX1 \label{tab:logtable}}
\tablewidth{0pt}
\tablehead{
\colhead{Serial No.} & \colhead{Observation ID} & \colhead{Date of Obs.} & \colhead{Epoch ID} &
 \colhead{Cleaned Exposure (ksec)}  \\
\nocolhead{} & \nocolhead{} &
\nocolhead{} & \nocolhead{} & \colhead{pn/MOS1/MOS2}
}
\startdata
1 & 824610101 & 2018-12-13 & XM1 & 71/89/94  \\
2 & 824610201 & 2018-12-19 & XM2 & 48/67/69  \\
3 & 824610301 & 2018-12-31 & XM3 & 50/66/70  \\
4 & 824610401 & 2019-01-02 & XM4 & 77/97/100 \\
\enddata
\tablecomments{The exposure times noted here are flare corrected approximate livetime CCD exposures.}
\end{deluxetable*}

\section{Data Analysis}\label{sec:data}
\xmm\ \citep{XMM2001} observed the NGC 4395 galaxy four times between December 2018 and January 2019 to study the AGN (Pi McHardy). 
ULX1 is around $\sim 3$ arcmin away from the galaxy's active nucleus. In the \xmm\ EPIC detectors, ULX1 is well isolated from any other X-ray sources. The details of the observation log are given in table \ref{tab:logtable}. The four observations studied here are abbreviated as XM1 (824610101), XM2 (824610201), XM3 (824610301), and XM4 (824610401) for ease of reference in the rest of the paper.

Using the standard data reduction procedure of \xmm\ data analysis software SAS v19.1.0 \footnote{\url{https://www.cosmos.esa.int/web/xmm-newton/sas}}, we clean the data from soft-proton and background flaring events and extract the science products for all EPIC-pn and MOS1/2 instruments. We select source photons from a circle of $25$ arcsec radius centered at $\alpha,\delta = 12:26:01.5,+33:31:31.0$ and background photons from a circle of $50$ arcsec radius in a nearby source-free region on the same chip. 

Unfortunately, ULX1 falls near the chip gap in all pn data. Additionally, the pn data are affected by strong bad column events in the source region in all observations. XM3 and XM4 data are mostly affected because a significant fraction of the source region falls in the chip gap, and the bad column passes through the central region of the source. This causes flux loss in pn spectra due to substantial charge loss. XM1 and XM2 pn data are comparatively less affected because the bad column passes through the edge of the source. We have performed spectral analysis of individual pn and MOS1/2 data for all epochs and found that the chip gap and bad column issue in pn observations do not affect the spectral profile except for a flux loss in XM3 and XM4 observations (see section \ref{sec:Spectral_Analysis} for details). Hence, for subsequent spectral analysis, pn and MOS1/2 data are fitted simultaneously for each observation with different spectral models.

The fast timing analysis is done on minimally-filtered data in order to maximize the number of counts and be sensitive to short-time variability. However, for the spectral analysis, we put a strict constraint of FLAG==0 to extract pn spectra for all observations to minimize the charge loss effect in the data. Spectra are grouped using \texttt{SPECGROUP} with a grouping factor of minimum $20$ counts per energy bin and oversampling factor $3$ for using $\chi^2$ statistics. Light curves are extracted using \texttt{EVSELECT} for single and double events in pn (PATTERN $<=$ 4) and singles, doubles, triples, and quadruples events in MOS (PATTERN $<=$ 12). The background-corrected source light curves are generated using the \texttt{EPICLCCORR} task, which corrects vignetting, bad pixels, chip gaps, PSF, and quantum efficiency. For timing analysis, we have performed barycentric correction on the events using \texttt{barycen} tool of SAS. Pileup in the data is evaluated with \texttt{EPATPLOT}, and no significant pileup is found.

We have also analyzed RGS data following standard data extraction procedure. However, the RGS spectra of the source are dominated by the background in all four observations. The combined RGS1+RGS2 spectral count rate (in full energy range) varies between $\sim 0.007-0.016$ counts/sec in different epochs, whereas the cleaned exposure varies between $\sim 171-226$ ksec.

\begin{figure*}
\centering
    \includegraphics[width=1. \linewidth]{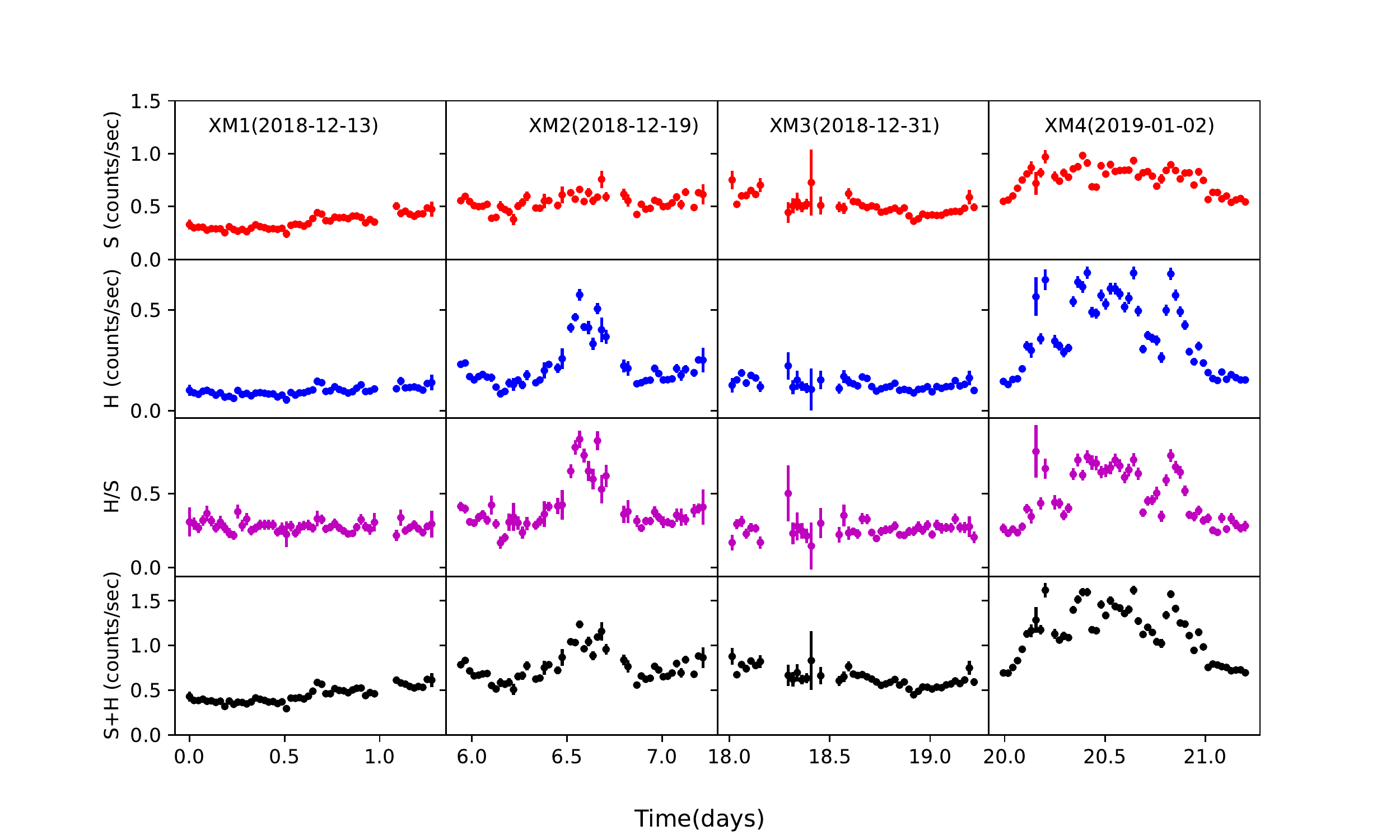}
\caption{\xmm\ pn light curves of ULX1 for four individual epochs binned by 2000 seconds. The first panel shows the soft ($0.3-1.0$ keV) count rate; the second panel shows the hard ($1.0-8.0$ keV) count rate, the third panel shows the hardness ratio defined as hard/soft photons count rate and the fourth panel shows the sum of soft and hard band lightcurve, i.e. the total lightcurve. First epoch (XM1) - shows a steady but slight uprising trend of the flux, Second Epoch (XM2) - A flaring episode occurred from around 30 ksec to 80 ksec of that observation, Third Epoch (XM3) - shows another steady but slightly declining trend of flux, Fourth Epoch (XM4) - finally, a large flaring episode happened covering most portion ($\sim 80$ ksec) of that observation. The prominent flares in XM2 and XM4 are predominant in the hard band, as seen in the hardness ratio plots.  
\label{fig:lightcurve}}
\end{figure*}

\section{Timing Analysis} \label{sec:Timing_Analysis}
Previous X-ray observations portrayed NGC 4395 ULX1 as the least variable source in a sample of soft ULXs studied in \citealt{Earnshaw2017}. However, the \xmm\ observations analyzed here show both short-term (in the time scale of a few kilo seconds) and long-term (in the time scale of a few days) timing variability.

The 2000-s binned \xmm\ pn lightcurves for all epochs are shown in figure \ref{fig:lightcurve}. From a visual inspection, it appears that while in XM1 and XM3, the source has a relatively steady flux, in XM2 and XM4, it shows flare-like activity. In order to check any energy-dependent nature of flaring behavior, we divide the time series into two energy bands. We set the soft band between 0.3 and 1\,keV (figure \ref{fig:lightcurve} first panel), and the hard band is above 1 keV (second panel). Although the background starts dominating above $\sim 5$ keV for XM1 and XM3 epochs and $\sim 8$ keV for XM2 and XM4 epochs (see Section~\ref{sec:ave_spec}), to make a direct comparison among all four epochs, we create the lightcurves in hard band between 1.0 and 8.0 keV for all four epochs. The hardness ratio in figure \ref{fig:lightcurve} third panel, is defined as the ratio of hard photon count rate to soft photon count rate. The fourth panel of the figure shows the sum of soft and hard band lightcurves. The figure shows how the ULX1 count rate varies between different epochs within a three week period. The XM2 observation shows a significant short-term flaring incident lasting $\sim 50$ ksec, mostly prominent in the hard energy band. The longest flaring episode is detected during the XM4 observation, where the flaring happened for $\sim 80$ ksec, a large portion of the observing span. The long flare of the XM4 epoch consists of multiple ephemeral sub-flaring episodes. However, the minimum count rate level of these transient sub-flares is much higher than the persistent count rate level of the XM1 and XM3 epochs. So, we consider the whole $\sim 80$ ksec long flare in XM4 as a single flaring epoch for our analysis. As there are no flaring events in XM1 and XM3, the hardness ratio is nearly constant for these two epochs. On the other hand, for XM2 and XM4, the hardness ratio demonstrates that the variability in different energy bands is prominent, and the flaring events are more significant in the harder spectral band (above $\sim 1$ keV).

\begin{figure*}
\includegraphics[width=0.49\textwidth, trim = {0 0cm 1.2cm 1.0cm}, clip]{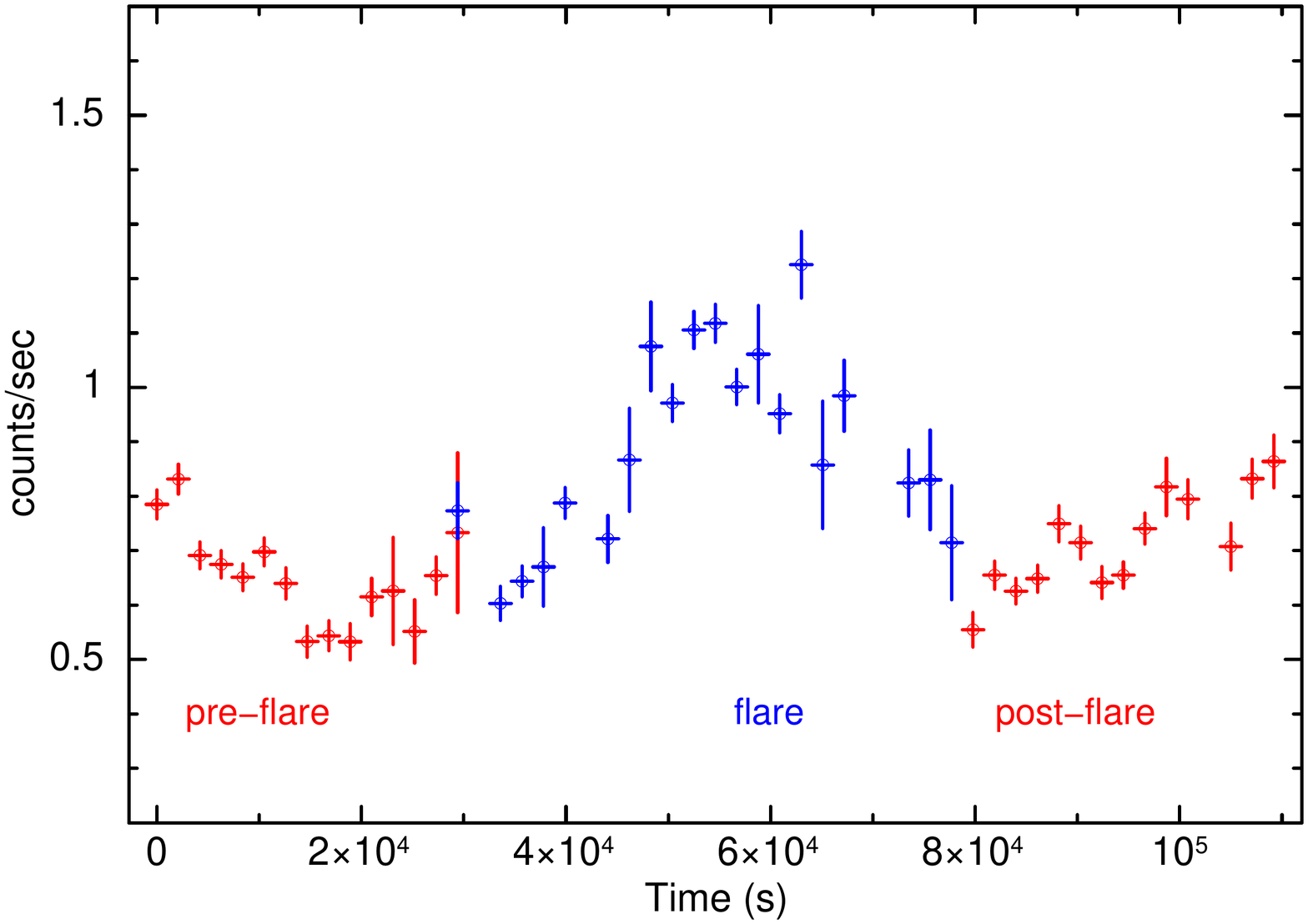}
    \includegraphics[width=0.49\textwidth, trim = {0 0cm 1.2cm 1.0cm}, clip]{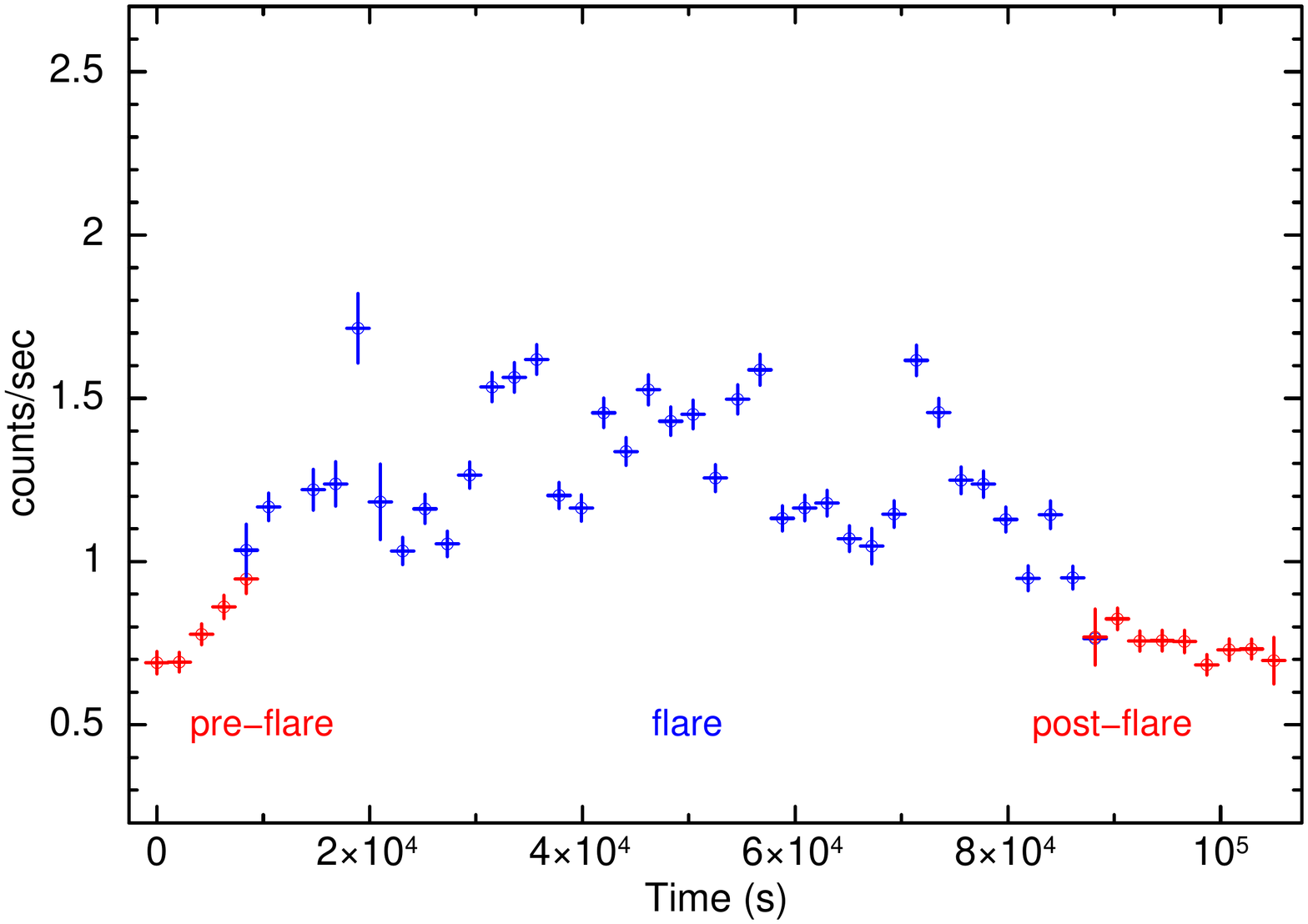}
\caption{Different transient intervals are defined for XM2 (left) and XM4 (right) epochs of observation. The pre-flare, flare, and post-flare intervals are separately indicated. The pre-flare and post-flare intervals are combined to get the ``non-flaring" interval.
\label{fig:flare_quies}}
\end{figure*}

Studying short-term variability is essential to decipher the inherent properties of transient events happening in ULXs. The first task to find such variability is to perform a Fourier space investigation of the time series. The power spectral density (PSD) analysis provides no evidence of quasi-periodic or periodic oscillation in any of the observations except for the presence of red noise at low frequencies in a few cases. We also search for transient pulsation in the time series. We incorporate the acceleration search technique to detect any transient pulse while correcting the Doppler shift due to binary orbital period correction. Tools like \texttt{HENDRICS} \citep{HENDRICS} and \texttt{PRESTO} \citep{PRESTO} are employed for these tasks. We utilize the EPIC-pn data for this purpose since it has the highest time resolution of $\sim 73.4$ ms. We use \texttt{HENACCELSEARCH} task of \texttt{HENDRICS} to search for pulsation in $0.3-8.0$ keV energy range and the frequency range of $0.01-6.8$ Hz to avoid artefact due to Nyquist limit. We use a maximum number of Fourier frequency bins ($z_{max}$) as 100 with a Fourier frequency bin resolution ($\Delta z$) of $1$. No significant pulsation is detected in any epoch. We also search for pulsation in the same energy and frequency ranges using \texttt{HENzsearch}, with a fast-folding algorithm which searches for the first spin derivative. Here also, no pulsation is found in any observation. However, this tool estimates an upper limit on the pulsed amplitude $\frac{I_{max}-I_{min}}{I_{max}+I_{min}}$ for the best candidate frequency within $90\%$ confidence, where $I_{max}$ and $I_{min}$ are the maximum and minimum values of the folded profile, respectively. We can estimate the upper limit on pulsed fraction $\frac{I_{max}-I_{min}}{I_{max}}$ from this pulsed amplitude value. We find that the upper limit of pulsed fraction varies between $\sim 10-17\%$ in these four epochs. We further divide XM2 and XM4 observations into three sub-epochs (see figure \ref{fig:flare_quies} left and right panel respectively): pre-flare, flare and post-flare intervals. Pre-flare and post-flare intervals have similar count rates and overlapping spectral properties. Hence they are combined and referred to as  ``non-flaring" interval. We perform the similar exercise to search for pulsation in those segmented ``flaring" and ``non-flaring" intervals using the aforementioned methods. No pulse period is found in either case. The upper limits of pulsed fraction for these segmented intervals vary between $\sim 11-23\%$. We also utilize the \texttt{accelsearch} tool from \texttt{PRESTO} package by invoking a ``jerk" search technique. Here, we use the maximum number of Fourier frequency bins 200 and the maximum number of Fourier frequency derivative bins $wmax = 600$ \citep{AndersenJerktech} within the same frequency range as before. Again, no significant pulsation is found.

The lightcurves of NGC 4395 ULX1 also demonstrate significant short-term variability in terms of the fractional root mean square (RMS) variability amplitude ($F_{var}$), which measures the variance of a source over the Poissonian noise in the time series, normalized to the average count rate \citep{Vaughan1,Edelson}. $F_{var} = \sqrt{\frac{S^2-\bar{{\sigma}^2}}{\bar{x}^2}}$, where $S^2 = \frac{1}{N-1}\sum_{i=1}^{N} (x_i - \bar{x})^2$ and $\bar{{\sigma}^2} = \frac{1}{N} \sum_{i=1}^{N} {\sigma_i}^2$. $x_i$ is the count rate at $i$'th bin, $\bar{x}$ is the mean count rate, $N$ is the total number of bins, $\sigma_i$ is the uncertainty in count rate in $i$'th bin. The error on $F_{var}$ is measured as $\sigma_{F_{var}} = \frac{1}{F_{var}} \sqrt{\frac{1}{2N}} \frac{S^2}{\bar{x}^2}$. We use the soft and hard lightcurves of each \xmm\ pn observation and bin them to $1000$ sec, then we estimate the $F_{var}$ and its error. The fractional variability values are listed in table \ref{tab:fvar}. We also utilize the MOS observations to verify whether the variability in pn lightcurves are an artefact of the bad column or chip gap as described in section~\ref{sec:data}. First we add MOS1 and MOS2 light curves to increase the count statistics (for both soft and hard energy bands) and bin the net light curves by $1000$ sec. We find that the same trend of variability is seen in both pn and MOS data (see table \ref{tab:fvar}). Hence, we confirm that these variabilities are intrinsic property of ULX1. This calculation clearly shows that fractional variability is higher in epochs XM2 and XM4 compared to the epochs XM1 and XM3 above $1.0$ keV.

\begin{deluxetable}{CC|C|C|C|}
\tablenum{2}
\tablecaption{Fractional variability in pn and MOS lightcurves for all four epochs.} \label{tab:fvar}
\tablewidth{0pt}
\tablehead{
\colhead{Epoch} & \multicolumn{2}{C}{pn} & \multicolumn{2}{C}{MOS} \\
\cline{2-3}\cline{4-5}
\colhead{} & \colhead{soft} & \colhead{hard} & \colhead{soft} & \colhead{hard}\\
}
\startdata
\hline
XM1 & 0.17 \pm 0.02 & 0.17 \pm 0.02 & 0.15 \pm 0.02 & 0.11 \pm 0.04  \\
XM2 & 0.11 \pm 0.01 & 0.52 \pm 0.04 & 0.12 \pm 0.02 & 0.47 \pm 0.04  \\
XM3 & 0.15 \pm 0.02 & 0.08 \pm 0.06 & 0.19 \pm 0.02 & 0.09 \pm 0.05  \\
XM4 & 0.15 \pm 0.01 & 0.47 \pm 0.03 & 0.14 \pm 0.01  & 0.47 \pm 0.03 \\
\enddata

\end{deluxetable}

\section{Spectral Analysis}\label{sec:Spectral_Analysis}
In this section, we report the spectral analysis results of the \xmm\ observations in detail. We use XSPECv12.12.0 \citep{XSPEC} for the spectral analysis throughout the paper. The absorption is quantified using the Tuebingen-Boulder ISM absorption model (\texttt{TBABS} in XSPEC) for Galactic and local extinction contributions. The updated abundance \citep{Wilms} and photoionization cross-section \citep{Verner} are used. The Galactic absorption column\footnote{\url{https://heasarc.gsfc.nasa.gov/cgi-bin/Tools/w3nh/w3nh.pl}} is kept fixed to $0.04 \times 10^{22}$ cm$^{-2}$ \citep{HI4PI}. The local absorption column is allowed to vary as a free parameter. Throughout the paper, the statistical uncertainties in the spectral parameters are within a 90\% confidence interval unless mentioned otherwise.

\begin{figure*}[!t]

\begin{minipage}[h]{1.0\linewidth}
\centering
\includegraphics[width=1.06\linewidth]{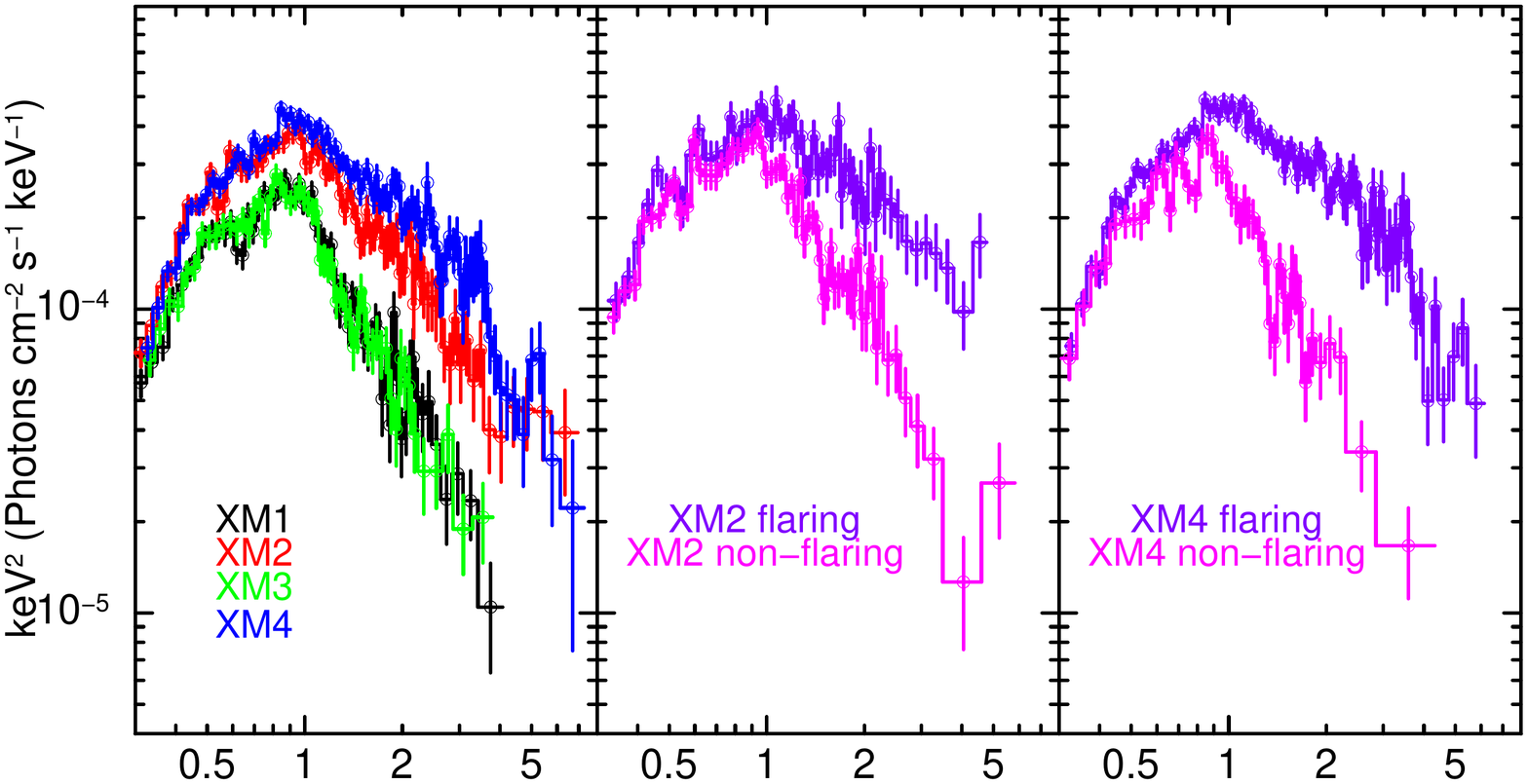}
\end{minipage}

\vspace{-7cm} 

\begin{minipage}[h]{1.0\linewidth}
\centering
\includegraphics[width=\linewidth]{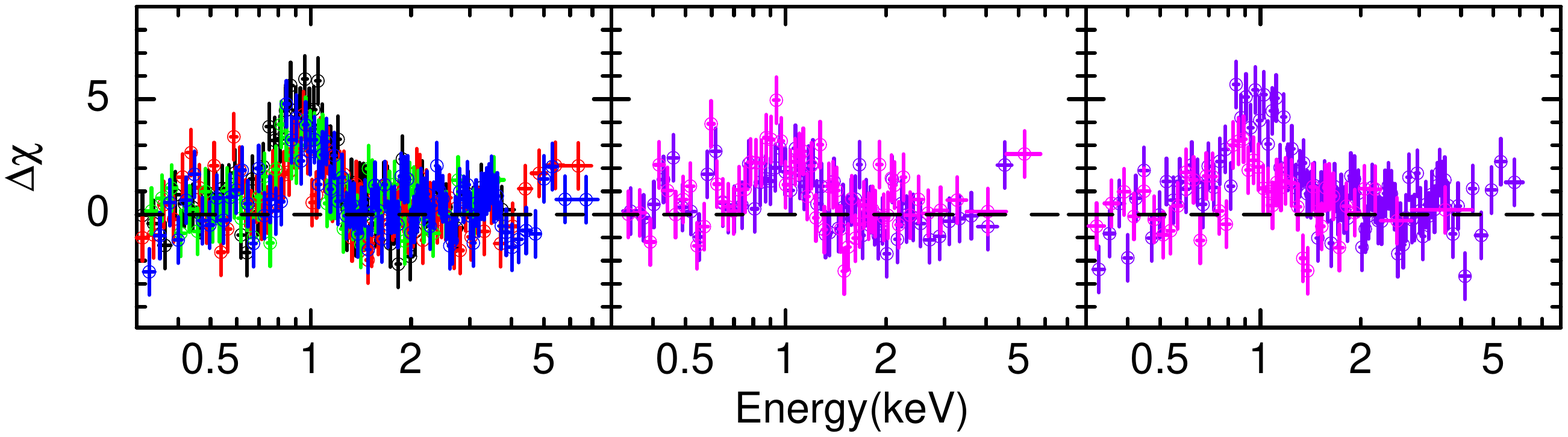}
\end{minipage}  

\vspace{-4cm}
\caption{Top: Unfolded MOS1 spectra for different epochs are plotted using \texttt{powerlaw} model of zero photon index ($\Gamma = 0$). For plotting purpose the normalization of the model is kept arbitrarily high. ULX1 exhibits a significant long-term spectral variability in different observation epochs. MOS1 spectra of XM1 and XM3 epochs in the left panel show overlapping spectral features. XM2 and XM4 epochs clearly show differences in spectral shapes and divergence in hard spectral regimes. Flaring and non-flaring spectra of XM2 and XM4 epochs are plotted in the middle and right panels. These two figures have a similar divergence in spectral characteristics above $1$ keV. Bottom: The residuals of the best fit continuum are shown for MOS1 spectra corresponding to the epochs shown in the top panels. To show the significant contribution of Gaussian in the spectra, for plotting purpose the Gaussian component is removed from the best fit model.}
\label{fig:spec_variab}
\end{figure*}

\subsection{Time averaged spectroscopy}\label{sec:ave_spec}
We start by analyzing the spectra of individual observations. We plot MOS1 spectra for all epochs (top left panel) in figure \ref{fig:spec_variab}. XM1 and XM3 have similar spectra in terms of flux and spectral feature, and both of them exhibit steep spectra extending only up to 5 keV, after which background starts dominating. XM2 and XM4 epochs have comparatively harder spectra and higher flux than the other two epochs and extend up to 8 keV, after which the signal-to-noise ratio decreases significantly.

Since there is a chip gap and strong bad column issue in pn data, we perform an exercise to verify whether these affect the source spectral properties. First of all, we carry out an individual analysis of pn and MOS1/2 data for all observations with a simple absorbed powerlaw model and a Gaussian component (see details of spectral models below) and find that the spectral parameters in the pn data are consistent with MOS1/2 data within 90\% statistical confidence interval, with the only exception of the normalization of the power law in XM3 and XM4.  Therefore, in the following, we simultaneously fit pn and MOS1/2 data for each observation with a cross-calibration constant fixed to 1 for MOS1 and left free to vary in MOS2 and pn. All other parameters of the models used are tied between the instruments. These cross-calibration values are within 10\% of MOS1, except for pn data of XM3 and XM4 epochs, the ones most affected by the noisy detector column and chip gap as described in Section~\ref{sec:data}.

To quantify the contribution of different emission mechanisms in ULX1 spectra, we fit them with various physical and phenomenological models in XSPEC. A simple powerlaw fit is useful to characterize the spectral hardness of the individual epochs.
XM1 and XM3 epochs have powerlaw index ($\Gamma$) value of $\sim 4.6$ and $\sim 4.8$ respectively, whereas XM2 and XM4 epochs have $\sim 3.7$ and $\sim 3.5$ respectively. It is interesting to compare the photon indices found for NGC 4395 ULX1 with those seen from canonical black hole X-ray binaries which can range from $\sim 1.7$ in low/hard state to $\sim 2.5$ in steep powerlaw state or very high state \citep{Remillard2006}. Thus, ULX1 has much steeper spectra in all four epochs compared to the sub-Eddington black hole X-ray binaries. Using this powerlaw continuum, a soft excess around $\sim 0.4$ keV and a Gaussian emission-like feature around $\sim 0.9$ keV are also detected. If we use a more complicated model composed of a thermal blackbody disk, a powerlaw, and a Gaussian, we are able to fit the ULX1 spectra for all epochs. However, the powerlaw model extends to low energies arbitrarily, making the parameters of the powerlaw model degenerate with those of the low energy components like the extinction or the emission lines. Hence, we explore other models, seeking a physically-consistent description of ULX1 spectra.

\begin{figure*}
\centering
    \includegraphics[width=1. \linewidth]{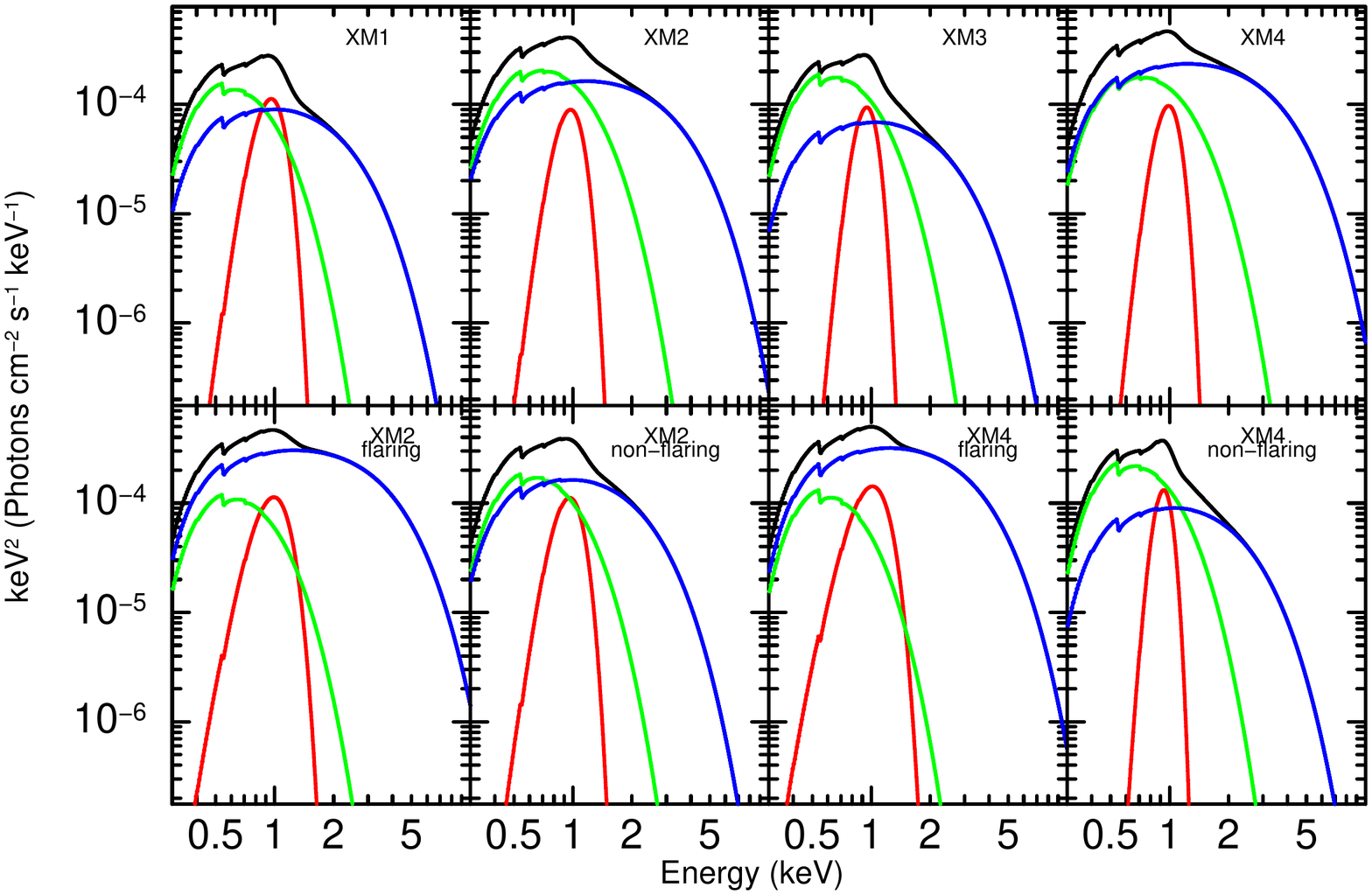}
\caption{The model components for MOS1 spectra for different epochs. The red component is the \texttt{GAUSSIAN}, green represents the \texttt{DISKBB} and blue represents \texttt{DISKPBB} components. Black represents the total model. The variation in \texttt{DISKPBB} model due to flaring incidents is clear from the figure .
\label{fig:eemo}}
\end{figure*}

Many ULX spectra, from soft ULXs like NGC 55 ULX and NGC 247 ULX-1 to moderately hard ULXs like NGC 1313 X1 and NGC 4559 X7, have been modeled with the composition of two thermal component models \citep{Pinto2017,Pinto2020,Kara2020,Walton2020, Pintore2021}. One thermal component describes the cooler accretion disk emission from the outer disk region, which is geometrically thin, and the other component describes the hot inner disk component, geometry of which depends on the accretion rate of the system. Especially for ULXs, when the accretion rate becomes close to or above the critical accretion rate, the disk becomes advection dominated disk or a slim accretion disk. This physical scenario motivates us to explore a continuum which consist of a thin accretion disk and a slim accretion disk. For NGC 4395 ULX1, we find that one cool thin disk (\texttt{DISKBB} in XSPEC) plus a comparatively hotter slim accretion disk (\texttt{DISKPBB} in XSPEC, with $p=0.50$) provide an adequate fit for the continuum at all epochs. We find that the parameter ``p'' of \texttt{DISKPBB} always assumes a value close to the lower limit of the parameter, $0.50$, corresponding to a slim disk regime. Hence we fix this ``p" value to $0.50$. In addition to the continuum, a broad Gaussian around $0.9$ keV is always required. We find that before addition of the Gaussian component, the $\chi^2/dof$ for only continuum fits are $402/171$ for XM1, $297/209$ for XM2, $255/143$ for XM3, and $375/241$ for XM4 epochs. These fits are significantly improved after the addition of a Gaussian component (see table \ref{tab:modelling}).  In XSPEC syntax, the best fit model we use is \texttt{TBABS(GAL)*TBABS*(GAUSS+DISKBB+DISKPBB)}. The $N_H$ value is consistent in all epochs ($\sim 0.05 \times 10^{22}$ cm$^{-2}$), and so is the Gaussian line energy ($\sim 0.9$ keV). The low-energy thin disk component (represented by the \texttt{DISKBB} model) also remains in a similar temperature state in all epochs. However, the \texttt{DISKPBB} component exhibits a higher temperature in the epochs XM2 and XM4 compared to the epochs XM1 and XM3. The average \texttt{DISKPBB} temperature in XM2 and XM4 epochs is $\sim 1.5$ times higher than that in XM1 and XM3 epochs. The best-fit parameters are noted in table \ref{tab:modelling} and the residuals from best fit model for MOS1 spectra are shown in bottom left panel of figure \ref{fig:spec_variab} for each observation. For visual purpose, the Gaussian component is removed from the best-fit model. Presence of prominent hump-like (or Gaussian) structure is evident from the residual plots. In figure \ref{fig:eemo}, (top panel), the model components are shown which depict the contribution of each component and their variation in different epochs.

\begin{deluxetable*}{CC|CC|CC}
\tablenum{3}
\tablecaption{Best fit model (\texttt{TBABS(GAL)*TBABS*(GAUSS+DISKBB+DISKPBB)}) parameters of NGC 4395 ULX1 for the four epochs. \label{tab:modelling}}
\tablewidth{0pt}
\tablehead{
\colhead{Parameter} & \colhead{Unit} & \colhead{XM1} & \colhead{XM3} &
\colhead{XM2} & \colhead{XM4} \\
}
\startdata
\hline
N_H & 10^{22} cm^{-2} & 0.05^{+0.03}_{-0.02} & 0.06^{+0.03}_{-0.02}  & 0.04 \pm 0.01  & 0.05 \pm 0.01 \\
E_{line} & keV & 0.91 \pm 0.02 & 0.92 \pm 0.02  & 0.92^{+0.03}_{-0.04} & 0.95^{+0.02}_{-0.03}  \\
\sigma  & keV & 0.15^{+0.03}_{-0.02} & 0.11^{+0.03}_{-0.02} & 0.14^{+0.04}_{-0.03} & 0.13 \pm 0.03 \\
norm & 10^{-5} & 5.71^{+2.90}_{-1.54} & 3.68^{+2.03}_{-0.96} & 4.12^{+2.44}_{-1.33} & 3.86^{+1.67}_{-1.03} \\
T_{thin} & keV & 0.17^{+0.03}_{-0.04} & 0.18^{+0.02}_{-0.03} & 0.22^{+0.02}_{-0.03} & 0.22^{+0.02}_{-0.03} \\
norm_{thin} &  & 55^{+95}_{-28} & 42^{+85}_{-18} & 20^{+19}_{-7} & 16^{+13}_{-6} \\
T_{slim} & keV & 0.54^{+0.07}_{-0.05} & 0.58^{+0.12}_{-0.09} & 0.78^{+0.08}_{-0.07} & 0.84 \pm 0.05 \\
norm_{slim} &  & 0.06^{+0.04}_{-0.03} & 0.03^{+0.06}_{-0.02} & 0.02 \pm 0.01 & 0.02 \pm 0.01 \\
\chi^2/dof &  & 165/168 & 150/140 & 218/206 & 275/238\\
F_x & 10^{-13} ergs/cm^2/s & 5.31 \pm 0.12 & 5.26 \pm 0.14  & 9.24 \pm 0.18 & 11.06 \pm 0.17\\
L_x & 10^{39} ergs/s & 1.44 \pm 0.03 & 1.43 \pm 0.04 & 2.51 \pm 0.05 & 3.0 \pm 0.05 \\
\enddata

\tablecomments{The absorbed flux $F_x$ and luminosity $L_x$ is calculated in 0.3-10.0 keV energy range. The Galactic absorption is fixed to $0.04 \times 10^{22}$ cm$^{-2}$. ``p" value of \texttt{DISKPBB} model is fixed to 0.50, resembling a slim disk. The distance is assumed to be 4.76 Mpc \citep{Vinokurov2016} to calculate the luminosity.}
\end{deluxetable*}

\subsection{Time-resolved spectroscopy}
ULX1 has undergone several flaring episodes, as described in section \ref{sec:Timing_Analysis}. This motivates us to carry out a comparative study of the source's spectral properties between flaring and non-flaring epochs. This subsection mainly focuses on epochs XM2 and XM4, where flares are prominent. As described earlier, we divide the XM2 and XM4 epochs into three sub-epochs: pre-flare, flare, and post-flare. Non-flaring intervals are chosen by adding the pre-flare and post-flare intervals because the spectral flux and properties overlap during these regimes. In figure \ref{fig:spec_variab} (top middle (XM2) and right (XM4) panels), we overplot the flaring and non-flaring spectra from MOS1. This figure shows how spectra of flaring and non-flaring states diverge mainly after $1$ keV (as in part already shown from the timing analysis). Below $1$ keV, the spectra are mostly consistent. 

We fit these time-resolved spectra with the same model used for the time-averaged analysis, since it was adequate to describe both the steady (XM1 and XM3) and the flaring (XM2 and XM4) observations. The fit of non-flaring and flaring spectra of individual observation is done simultaneously, linking absorption and line energy and letting the disk parameters vary freely. We report the best fit parameter and error estimates in table \ref{tab:modelling2}. Figure \ref{fig:spec_variab} (bottom middle (XM2) and right (XM4) panels) shows the residuals from best-fit model (without the Gaussian component) for MOS1 spectra. The non-varying nature of \texttt{DISKBB} temperature between flaring and non-flaring episodes is similar to the case of time-averaged spectroscopic results reported in table \ref{tab:modelling}. The best fit \texttt{DISKPBB} temperatures in flaring episodes are higher than that in non-flaring episodes of XM2 and XM4 by $\sim 1.7$ and $\sim 1.4$ times, respectively. Also, as expected, the spectral parameters of the non-flaring episodes of XM2 and XM4 are consistent with the parameters of XM1 and XM3 epochs which do not show any flaring events. Especially, the temperatures of the thin and slim disks are similar in all of these cases which suggest a steady accretion in the system during these steady (XM1 and XM3) epochs and non-flaring episodes of XM2 and XM4. In figure \ref{fig:eemo}, (bottom panel), the model components for flaring and non-flaring spectra are shown which manifest how spectral components vary due to flaring events, specifically the \texttt{DISKPBB} component.

\begin{deluxetable*}{CC|CC|CC}
\tablenum{4}
\tablecaption{Time resolved spectral parameters of XM2 and XM4 epochs using best fit model same as in table \ref{tab:modelling}. \label{tab:modelling2}}
\tablewidth{0pt}
\tablehead{
\colhead{Parameter} & \colhead{Unit} & \multicolumn{2}{C}{XM2} & \multicolumn{2}{C}{XM4} \\
\cline{3-4}\cline{5-6}
\colhead{} & \colhead{} & \colhead{flaring} & \colhead{non-flaring} &
\colhead{flaring} & \colhead{non-flaring} \\
}
\startdata
\hline
N_H & 10^{22} cm^{-2} & 0.05^{+0.07}_{-0.02} &   & 0.07 \pm 0.02  &  \\
E_{line} & keV & 0.90^{+0.03}_{-0.02} &   & 0.91^{+0.02}_{-0.04} &   \\
\sigma  & keV & 0.19^{+0.11}_{-0.08} & 0.15^{+0.12}_{-0.05} & 0.21^{+0.02}_{-0.03} & 0.09^{+0.06}_{-0.03} \\
norm & 10^{-5} & 7.09^{+10.98}_{-4.65} & 5.83^{+29.76}_{-2.5} & 9.63^{+2.87}_{-3.85} & 4.35^{+5.33}_{-1.41} \\
T_{thin} & keV & 0.18^{+0.10}_{-0.09} & 0.18^{+0.04}_{-0.09} & 0.15^{+0.07}_{-0.02} & 0.18^{+0.02}_{-0.05} \\
norm_{thin} &  & 32^{+3983}_{-27} & 41^{+3139}_{-23} & 68^{+259}_{-59} & 58^{+204}_{-26} \\
T_{slim} & keV & 0.89^{+0.21}_{-0.08} & 0.52^{+0.08}_{-0.05} & 0.80^{+0.02}_{-0.03} & 0.56^{+0.18}_{-0.11} \\
norm_{slim} &  & 0.02 \pm 0.01 & 0.11^{+0.12}_{-0.07} & 0.04 \pm 0.01 & 0.05^{+0.14}_{-0.04} \\
\chi^2/dof &  & 349/327 &  & 409/344 & \\
F_x & 10^{-13} ergs/cm^2/s & 12.36^{+0.45}_{-0.42} & 7.87 \pm 0.21  & 12.49 \pm 0.21 & 6.50 \pm 0.25 \\
L_x & 10^{39} ergs/s & 3.35^{+0.12}_{-0.11} & 2.14 \pm 0.06 & 3.39 \pm 0.06 & 1.76 \pm 0.07 \\
\enddata

\tablecomments{The absorbed flux $F_x$ and luminosity $L_x$ is calculated in 0.3-10.0 keV energy range. The Galactic absorption is fixed to $0.04 \times 10^{22}$ cm$^{-2}$. ``p" value of \texttt{DISKPBB} model is fixed to 0.50, resembling a slim disk. The distance is assumed to be 4.76 Mpc \citep{Vinokurov2016} to calculate the luminosity.}
\end{deluxetable*}

\begin{figure}
\centering
    \includegraphics[width=1. \linewidth]{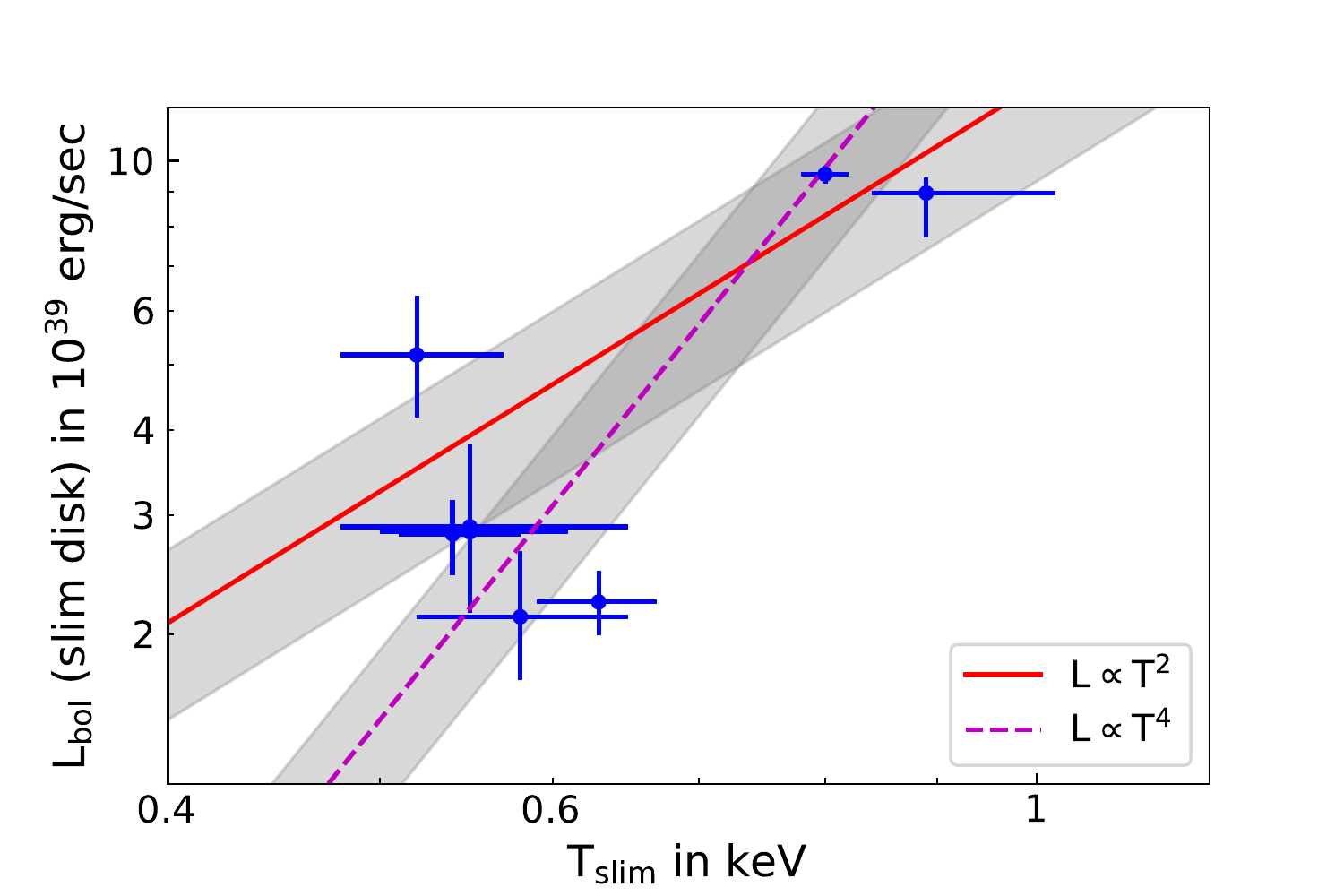}
\caption{Relation between unabsorbed bolometric luminosity of the hard slim disk component vs. temperature of that disk. The shaded regions show $95\%$ confidence intervals on the normalizations of the powerlaw relations.
\label{fig:T_L}}
\end{figure}

\section{Discussion and Conclusions} \label{sec:Discussion}
A previous study by \citealt{Earnshaw2017} showed that the spectra from ULX1 exhibit a steep powerlaw tail and a strong, Gaussian-like feature near $\sim 0.9$ keV, which could be explained by a mekal model. However, multiple studies suggest that two-component disks might be more appropriate to describe spectra of ULXs \citep{Pinto2017,Pinto2020,Kara2020,Walton2020}. 

It is widely believed that disks around ULXs have a two-tiered structure: very far from the ULX, the disk is a typical \citealt{Shakura1973}-like thin disk. Once the local luminosity of the disk approaches the Eddington limit, the disk ``inflates'', winds are launched, carrying away excess mass, and advection plays a central role. In this physical scenario, the low-energy thin disk component corresponds to the outer disk, and the high-energy slim disk component corresponds to the emission from the inner disk and the winds. By setting this physical accretion picture, we will discuss how these new \xmm\ observations conform with these scenarios.
 
The \xmm\ observations of NGC 4395 ULX1 before 2018 do not show significant short-term variability in the source \citep{Earnshaw2017}. The latest 2018-2019 observations, however, show significant flaring episodes in the data. We compare the ULX1 spectral properties with known ULX categories. The hard spectra generated from the radial advection of slim accretion disk modeled by a \texttt{DISKPBB} component implies super-critical accretion scenario. The luminosity measurement ($L \sim 1.4-3.0 \times 10^{39} \lumcgs$) shows that ULX1 is accreting at or just above the Eddington accretion rate limit if it hosts a typical $\sim 10 \ms$ black hole. Thus, one might initially expect ULX1 to exhibit a broadened disk spectrum. However, it is crucial to note that the temperature of hot slim disk component is less than typical temperature of BD ULXs. Also, ULX1 spectra require cool disk component which is not usual for BD ULXs. Alternatively, one might compare ULX1 with SULs and ULSs judging from the soft nature of the source. In all observations, the source exhibits a powerlaw photon index of $\Gamma > 2$ in $0.3-5.0$ keV band, hence apparent classification would be an SUL. However, it is important to consider that many SUL sources \citep{Sutton2013} show higher luminosity than the luminosity range of NGC 4395 ULX1. While comparing with ULSs, we observe that the two disk component spectra in ULX1 are quite distinctive in nature from the single ultrasoft blackbody spectra of ULSs. Thus, we suggest that NGC 4395 ULX1 is a case that is intermediate between SUL and ULS states. The spectral properties of NGC 4395 ULX1, like the spectral profile, two thermal component continuum, and atomic emission features, are similar to those of NGC 55 ULX and NGC 247 ULX-1, in particular \citep{Pinto2017,Pinto2021}. One possible explanation for the soft nature of the source is a geometrical picture when the line of sight is nearer to the plane of the disk, thus obstructing the hard photons coming from the inner and hotter region of the accretion disk (see similar discussion for NGC 55 ULX in \citealt{Pinto2017}).

The power spectrum analysis of the time series shows low-frequency red noise explained by power-law type PSD in a few observations. This often occurs in X-ray binaries due to variation in mass accretion rate \citep{Uttley}. We do not find any quasi-periodic or periodic oscillation on top of the red and white noise in the power spectrum. The soft and hard energy time series in XM2 and XM4 epochs confirm that the harder spectral components dominate the flaring activity. A comparable behavior in other ULXs \citep[e.g.][]{Middleton2015Mar,Gurpide2021a,Gurpide2021b} can be interpreted as the partial occultation of the inner region of the disk from the wind launched by the super-Eddington disk. This might in principle be the explanation for our results. Another possibility is that we are witnessing a genuine change in accretion rate, leading to the flaring behavior.

An advection-dominated disk, in the absence of beaming, is expected to show a $L \propto T^2$ relation \citep{Kubota2004}. This might very well be the case in our data, as shown in figure \ref{fig:T_L}, where we plot the unabsorbed bolometric luminosity ($0.01-10.0$ keV) from \texttt{DISKPBB} component and the corresponding disk temperature. To obtain the bolometric \texttt{DISKPBB} flux ($F_{bol}$) we have extended the energies of the instrumental responses in XSPEC following a similar study by \citealt{Urquhart2016}. To increase the statistics for luminosity vs. temperature plot, we have also included two additional \xmm\ observations (ID - 0142830101,0200340101), which had comparable exposures to that of the four observations studied here (see appendix \ref{Appendix1} for the spectral analysis results of these two observations). During these two observations, the source exhibits a similar spectral properties to that of the non-flaring epochs of the new observations when fitted with same spectral model. We obtain the bolometric luminosity and disk temperature and their $1\sigma$ errors from all observations and fit the data (using scipy.odr routine \footnote{\url{https://docs.scipy.org/doc/scipy/reference/odr.html}}; \citealt{Boggs1990}) with a $L\propto T^2$ and a $L\propto T^4$ relation. In pursuance of adopting a conservative approach, we consider the larger error on both axes for fitting the data points. Figure \ref{fig:T_L} shows that the luminosity-temperature plane of \texttt{DISKPBB} model is broadly consistent with both $L\propto T^2$ and $L\propto T^4$ curves. The shaded regions in the figure show $95\%$ confidence intervals on the normalizations of the powerlaw relations.

Typically bolometric luminosity would be $L_{bol} = \frac{2\pi D^2}{\cos\theta} F_{bol}$, where $\theta$ is the disk inclination and $D$ is the distance to the source. A standard assumption is that the disk inclination of these sources is $\sim 60^{\circ}$ (see \citealt{Urquhart2016} for details), hence $L_{bol}=4\pi D^2F_{bol}$, which has been used while estimating the luminosity. Since the inclination angle of the accretion disk is highly uncertain, the absolute value of bolometric luminosity should be taken with caution. However, for a fixed inclination in all epochs, the luminosity-temperature plane would exhibit a fixed positive powerlaw relation. It is important to note that, although the data broadly follows both the curves, empirically $L\propto T^2$ relation is more appropriate for an advection dominated accretion (or slim) disk.

The accretion rate of a slim disk in the presence of advection, relates to the luminosity as $L \sim L_{edd}[1+\ln\dot{m}]$, where $\dot{m}$ is the Eddington factor, ratio of the accretion rate to the Eddington accretion rate \citep{Shakura1973}. The ratio of the luminosity during flares to the non-flare periods ($\frac{L_{flare}}{L_{non-flare}} \sim 2$) can then be used to estimate the change of mass accretion rate. Simple algebra leads to the relation of corresponding Eddington factors of $\dot{m}_{flare} \simeq e\cdot \dot{m}^2_{non-flare}$. 

Some authors propose that an apparent change of luminosity in super-Eddington disks might be dominated by the geometrical beaming from the disk winds. In this case, $L \propto L_{edd}[1+\ln\dot{m}]\dot{m}^2$, because beaming factor has been proposed to scale
as $73/\dot{m}^2$ \citep{King2009,King2016,King2017}. Since the luminosity ratio in this case is small ($\sim 2$), the luminosity-accretion rate relation can be approximated to $\frac{L_{flare}}{L_{non-flare}} \sim \frac{\dot{m}^2_{flare}}{\dot{m}^2_{non-flare}}$. In that scenario, the Eddington factors corresponding to flaring and non-flaring episodes follow a simple form of $\dot{m}_{flare} \simeq \sqrt{2} \dot{m}_{non-flare}$.

The accretion disk normalizations provide an estimate for the inner radius $R_{in} \simeq \xi \kappa^2 N^{\frac{1}{2}} d_{10} (\cos\theta)^{-\frac{1}{2}}$ km, where $\xi$ is the geometric correction factor and $\kappa$ is the color correction factor, $d_{10}$ is the distance in 10 kpc unit, N is the normalization and $\theta$ is the inclination angle of the disk \citep{Kubota1998, Soria2015}. We estimate the inner radius from the disk normalizations for all the four epochs XM1, XM2, XM3 and XM4. We understand that due to large uncertainty in the \texttt{DISKBB} normalization measurement in different epochs, the thin accretion disk radius can have a large range of values between $\sim 2000 (\cos\theta)^{-\frac{1}{2}}$ km to $\sim 7000  (\cos\theta)^{-\frac{1}{2}}$ km assuming $\xi \sim 0.412$ and $\kappa \sim 1.7$ \citep{Shimura1995,Kubota1998}. The \texttt{DISKPBB} slim accretion disk normalizations are however similar in different epochs, hence the inner radius is found by taking a simple average of \texttt{DISKPBB} normalization from all the four epochs. We would caution that, to estimate such inner radius from \texttt{DISKPBB} model, we assume that the radius is constant which is appropriate for $L \propto T^4$ relation. However, from $L \propto T^2$ relation, the inner radius need not necessarily be constant unless advection plays a significant role. Also it should be noted that an estimate of radius from \texttt{DISKPBB} model is an approximation of the advection dominated disk since the \texttt{DISKPBB} is an approximate powerlaw scaled model of the radial dependent temperature and does not formally include the physical effects of advection on the inferred inner radius.

Under the assumption of $\xi \sim 0.353$ and $\kappa \sim 3$ for a slim disk \citep{Vierdayanti, Soria2015}, the radius turns out to be $\sim 273  (\cos\theta)^{-\frac{1}{2}}$ km. This would correspond to the last stable orbit ($R_{ISCO} = 6GM/c^2$) of a $\sim 31$ \ms non-rotating black hole assuming a face on disk or a $\sim 43$ \ms non-rotating black hole if the disk inclination is $\sim 60^{\circ}$. On the other hand, this radius would correspond to the magnetospheric radius ($R_M = 7 \times 10^7 \Lambda m^{\frac{1}{7}} R_6^{\frac{10}{7}} B_{12}^{\frac{4}{7}} L_{39}^{\frac{-2}{7}}$ cm; parameters are explained in \citealt{Mushtukov2017}) of a magnetized neutron star of $1.4$ \ms with a magnetic field of $\sim 6.1 \times 10^{11}$ Gauss for face on disk geometry or $\sim 1.58 \times 10^{12}$ Gauss for disk inclination of $\sim 60^{\circ}$. We have used the typical values of $\Lambda \sim 0.5$, neutron star radius of $ \sim 10^6$ cm, and the average luminosity (in $0.3-10.0$ keV) measured for the four epochs $\sim 2.1 \times 10^{39}$ \lumcgs for disk inclination of $\sim 60^{\circ}$ or two times less for face on disk.

If the above interpretations are true, it is the inner disk component that is dominating the variable part of the spectrum. This might be due either to an intrinsic change of accretion rate, or to a variable clumpy wind that partially occults the inner region and imprints this variability on the hard emission. However, the measured neutral absorption column density does not change during these observations. This suggests that if the wind clouds which block the hotter portions of the disk are not highly ionized, such transient flaring phenomena are related to the inner disk region far distanced from the wind cloud regions. However, if those regions of wind clouds are highly ionized, changes in line-of-sight scattering would imprint such variability.

Another plausible inference regarding hard photons observed during flares could be related to the inverse-Compton scattering process. Due to flaring events, the number of inner disk photons can proliferate owing to the high accretion rate. Those photons can further interact with high energetic electrons via the inverse-Compton process and release harder photons from the Coronal region. Thus producing spectrally harder flaring events compared to non-flaring events.

NGC 4395 ULX1 shows a clear broad emission feature around $\sim 0.9$ keV. Similar feature has been reported in several other ULXs like NGC 1313 X1, NGC 55 ULX, NGC 247 ULX-1 \citep{Pinto2017,Pinto2020,Pinto2021}. This broad $\sim 0.9$ keV line feature is in reality a combination of multiple emission lines which cannot be resolved by EPIC instruments. These lines are typically associated with Mg XII, Fe XXII-XXIII, Ne X, Ne IX, O VIII, and O VII lines as also observed in soft ULXs like NGC 55 ULX or NGC 247 ULX-1 (see e.g., \citealt{Pinto2017, Kosec2021}). In fact, the broad feature around $\sim 1$ keV in EPIC data can be modelled by emission lines around $\sim 0.9$ keV or absorptions around $\sim 0.7$ and $\sim 1.2$ keV \citep{Middleton2014,Middleton2015Mar}. In case of EPIC+RGS combined data, one can fit all emission and absorption features to explain the broad feature (see \citealt{Pinto2021}). However, with low spectral resolution EPIC only data, it is a custom to either fit a Gaussian emission or two absorptions (\texttt{GABS}) (see \citealt{Middleton2015Mar} for details). Due to well constrained parameters and simpler nature of \texttt{GAUSS} model fit in our data, we utilize this model to explain the broad hump-like feature. However, we cannot discard the presence of absorption lines in ULX1 spectra within the limited spectral resolution of EPIC data. The signal-to-noise ratio in RGS data is poor for ULX1 in all four observations, as the background mostly dominates the whole RGS spectra, hence, we could not use RGS data to quantify any emission or absorption feature present in the source. Future X-ray monitoring of the source will be crucial in establishing its transient nature and better understanding of the physical properties.


%


\begin{acknowledgments}
Authors would like to thank the anonymous referee for the positive comments and useful suggestions which have helped improving the manuscript significantly. The scientific results reported in this article have used archival data (available at the High Energy Astrophysics Science Archive Research Center (HEASARC)) obtained with \xmm, an ESA science mission with instruments and contributions directly funded by ESA member states and NASA. We would like to thank the HEASARC and XMM-Newton helpdesk team members for their valuable support.
\end{acknowledgments}

\vspace{5mm}
\facilities{\xmm; \citet{XMM2001}}

\software{ HEASOFT (\url{https://heasarc.gsfc.nasa.gov/docs/software/heasoft/}; \citet{Heasoft2014}), FTOOLS (\url{https://heasarc.gsfc.nasa.gov/ftools/}); \citet{Ftools1995,Ftools1999}, \xmm\ SAS (\url{https://www.cosmos.esa.int/web/xmm-newton/sas}; \citet{Gabriel2004}), HENDRICS (\url{https://hendrics.stingray.science/en/latest/}; \citet{HENDRICS}), PRESTO (\url{https://github.com/scottransom/presto}; \citet{PRESTO})
}



\appendix
\section{Analysis results of \xmm\ observations (ID - 0142830101 and 0200340101)}\label{Appendix1}
We discuss the spectral analysis results of ULX1 for the observations 0142830101 (Date: 2003-11-30) and 0200340101 (Date: 2004-06-02). The data reduction process followed for these two observations is similar to that described in section \ref{sec:data}. In both observations, the source exhibits steep spectra extending only up to 5 keV, similar to XM1 and XM3 epochs, after which background starts dominating. The pn observation of 0142830101 is affected by chip gap and bad column, as happened in other observations discussed in the paper. We have given a similar treatment as discussed in sections \ref{sec:data} and \ref{sec:Spectral_Analysis}, to verify the effect of the chip gap and bad column on source spectral properties in this observation. We simultaneously fit pn and MOS1/2 data by keeping cross-calibration constant fixed to 1 for MOS1 and left free to vary for MOS2 and pn. In observation 0200340101, the source is highly off-axis (see also \citealt{Earnshaw2017}) and falls out of the pn detector field of view. Only MOS1 and MOS2 data are utilized for this observation. The  cleaned exposures of pn/MOS1/MOS2 for 0142830101 observation are $69/93/94$ ksec. For 0200340101 observation, the MOS1/MOS2 cleaned exposures are $65/65$ ksec. The spectral analysis results with the model \texttt{TBABS(GAL)*TBABS*(GAUSS+DISKBB+DISKPBB)} are tabulated below.


\begin{deluxetable*}{CC|CC}[h]
\tablenum{A1}
\tablecaption{Best fit model (\texttt{TBABS(GAL)*TBABS*(GAUSS+DISKBB+DISKPBB)}) parameters of NGC 4395 ULX1 for observations 0142830101 and 0200340101. \label{tab:archival_observation_results}}
\tablewidth{0pt}
\tablehead{
\colhead{Parameter} & \colhead{Unit} & \colhead{0142830101} & \colhead{0200340101} \\
}
\startdata
\hline
N_H & 10^{22} cm^{-2} & 0.06 \pm 0.02 & 0.12^{+0.08}_{-0.06} \\
E_{line} & keV & 0.94 \pm 0.02 & 0.92^{+0.04}_{-0.06}  \\
\sigma  & keV & 0.09 \pm 0.02 & 0.13 \pm 0.03 \\
norm & 10^{-5} & 2.37^{+0.71}_{-0.50} & 6.11^{+5.75}_{-2.47} \\
T_{thin} & keV & 0.18 \pm 0.02 & 0.13^{+0.05}_{-0.04} \\
norm_{thin} &  & 27^{+35}_{-13} & 216^{+2573}_{-184} \\
T_{slim} & keV & 0.63^{+0.08}_{-0.06} & 0.55^{+0.11}_{-0.08} \\
norm_{slim} &  & 0.02^{+0.02}_{-0.01} & 0.05^{+0.06}_{-0.03}\\
\chi^2/dof &  & 158/155 & 94/75\\
F_x & 10^{-13} ergs/cm^2/s & 4.07^{+0.11}_{-0.10} & 4.19 \pm 0.21\\
L_x & 10^{39} ergs/s & 1.10 \pm 0.03 & 1.14 \pm 0.06\\
\enddata

\tablecomments{The absorbed flux $F_x$ and luminosity $L_x$ is calculated in 0.3-10.0 keV energy range. The Galactic absorption is fixed to $0.04 \times 10^{22}$ cm$^{-2}$. ``p" value of \texttt{DISKPBB} model is fixed to 0.50, resembling a slim disk. The distance is assumed to be 4.76 Mpc \citep{Vinokurov2016} to calculate the luminosity.}
\end{deluxetable*}

\bibliography{NGC4395ULX1}{}
\bibliographystyle{aasjournal}



\end{document}